\documentclass[a4paper,12pt]{article}
\usepackage{graphics}
\def\beq{\begin{equation}}
\def\eeq{\end{equation}}
\def\bea{\begin{eqnarray}}
\def\eea{\end{eqnarray}}
\def\half{\frac{1}{2}}
\def\ve{{\varepsilon}}
\def\cop{\alpha}
\begin{document}
\title{An Equivalent Hermitian Hamiltonian for the non-Hermitian $-x^4$ Potential}
\author{H.~F.~Jones\\ Physics Department, Imperial College, London SW7, UK \and
J.~Mateo \\ Departamento de F\'isica Te\'orica, Facultad de
Ciencias, \\E-47011, Valladolid, Spain}
\date{}
\maketitle

\begin{abstract}
The potential $V(x)=-x^4$, which is unbounded below on the real
line, can give rise to a well-posed bound state problem when $x$
is taken on a contour in the lower-half complex plane. It is then
$PT$-symmetric rather than Hermitian. Nonetheless it has been
shown numerically to have a real spectrum, and a proof of reality,
involving the correspondence between ordinary differential
equations and integrable systems, was subsequently constructed for
the general class of potentials $-(ix)^N$. For such Hamiltonians
the natural $PT$ metric is not positive definite, but a
dynamically-defined positive-definite metric can be defined,
depending on an operator $Q$. Further, with the help of this
operator an equivalent Hermitian Hamiltonian $h$ can be
constructed. This programme has been carried out exactly for a few
soluble models, and the first few terms of a perturbative
expansion have been found for the potential $m^2x^2+igx^3$.
However, until now, the $-x^4$ potential has proved intractable.
In the present paper we give explicit, closed-form expressions for
$Q$ and $h$, which are made possible by a particular
parametrization of the contour in the complex plane on which the
problem is defined. This constitutes an explicit proof of the
reality of the spectrum. The resulting equivalent Hamiltonian has
a potential with a positive quartic term together with a linear
term.
\end{abstract}

\section{Introduction}

There has been a great deal of interest in non-Hermitian
Hamiltonians since the numerical observation by Bender and
Boettcher~\cite{BB} that Hamiltonians of the form \beq\label{HBB}
H=p^2-g (ix)^N \eeq have a real positive spectrum for $N\ge 2$. As
illustrated in Fig.~1 (from Ref.~\cite{BB}, where $g=1$), their
spectra constitute a smooth extrapolation from the simple harmonic
oscillator, for which $N= 2$. The reality of their spectra is
understood as being due to their unbroken $PT$ symmetry, but there
is no simple way of telling in advance whether or not this
symmetry is broken, as indeed it is for $N < 2$, where the spectra
are partly complex. Eventually a rather intricate proof of the
reality of the spectrum, involving the correspondence between the
differential equations for such potentials and integrable models,
was constructed by Dorey et al.~\cite{Dorey}.
\begin{figure}[h]
\hspace{2cm}\resizebox{!}{3.6in}{\includegraphics{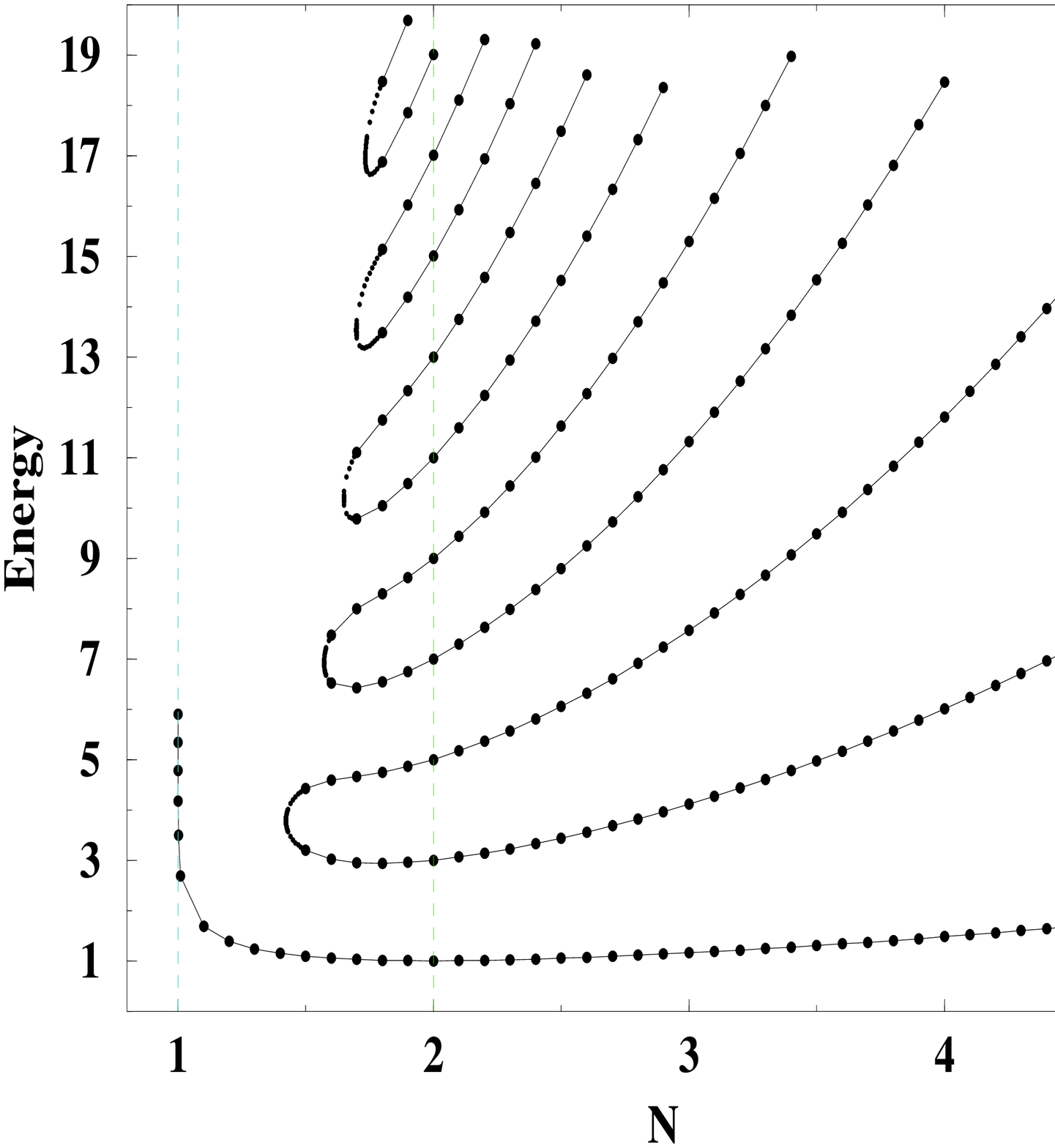}}
\caption{Energy levels of Eq.~(\ref{HBB}) for $N=4$ with $g=1$, from Ref.~\cite{BB}}
\end{figure}

A potential problem with such Hamiltonians is their physical
interpretation, since the natural $PT$ norm on the Hilbert space,
$\int dx \psi^*(x)\psi(-x)$, is not positive definite, in contrast
to the usual norm $\int dx \psi^*(x)\psi(x)$. However, it turns
out to be possible to construct an alternative norm, the $CPT$
norm~\cite{PCT}, which is indeed positive definite. This norm is
different from the usual norm, in that it is dynamically
determined by the Hamiltonian itself, and needs to be calculated
in each individual case.

Such calculations were encompassed by Mostafazadeh~\cite{AM} in the
more general framework of pseudo-Hermiticity, whereby
\bea\label{Hdag}
H^\dag=\eta H \eta^{-1}\,
\eea
Here the operator
$\eta$ is Hermitian and positive definite, and may usefully be
written as $\eta=e^{-Q}$, in order to connect with the notation of
Ref.~\cite{Q}, where, for $PT$-symmetric Hamiltonians,
$\eta=PC$ and $Q$ was defined by $C=e^Q P$.
For calculational purposes it is much easier to deal with
$Q$ rather than $\eta$ directly. Mostafazadeh showed further that
\bea\label{h} h\equiv e^{-\half Q} H e^{\half Q} \eea is an
equivalent Hermitian Hamiltonian, obtained from $H$ by a
similarity (Darboux) transformation.

In general it is difficult to solve Eqs.~(\ref{Hdag}), (\ref{h})
exactly; instead one uses perturbation theory in a small parameter
$\ve$. If $H$ is of the form $H=H_0+\ve H_1$, where $H_0$ is
Hermitian and $H_1$ anti-Hermitian, then $Q$ can be taken as
$Q=\sum_{r\, {\rm odd}} Q_r \ve^r$, which then gives $h=\sum_{r\, {\rm
even}} h_r \ve^r$. In this case
the first few equations for the $Q_r$, arising from the expansion
of Eq.~(\ref{Hdag}), read\footnote{The equations of even order are
satisfied identically by $Q_{2n}=0$.}
\bea\label{Qs}
\left[Q_1,H_0\right]&=& 2H_1\\
\left[Q_3,H_0\right]&=& {\textstyle\frac{1}{6}}[Q_1,[Q_1,H_1]]\cr
\left[Q_5,H_0\right]&=&\rule{0cm}{.5cm}
{\textstyle\frac{1}{6}}([Q_3,[Q_1,H_1]]+[Q_1,[Q_3,H_1]])
-{\textstyle\frac{1}{360}}[Q_1,[Q_1,[Q_1,[Q_1,H_1]]]]\nonumber
\eea
and so on. Using these, the first few equations for the $h_r$, arising
from the expansion of Eq.~(\ref{h}), can be cast in the form
\bea\label{hs}
h_0&=&H_0\cr
h_2&=&-{\textstyle\frac{1}{4}}[Q_1,H_1]\\
h_4&=&{\textstyle\frac{1}{192}}[Q_1,[Q_1,[Q_1,H_1]]]-{\textstyle\frac{1}{4}}[Q_3,H_1]\nonumber\,.
\eea

The smooth continuation from the harmonic oscillator, and the
ODE-IM correspondence of Ref.~\cite{Dorey}, rest on the fact that
the Schr\"odinger differential equation has several different
sectors, defined by wedges in the complex $x$-plane. Along the
centre of the wedges the wave-function decays exponentially at
infinity, while along the edges the wave-function is purely
oscillatory. Figure 2, taken from Ref.~\cite{BB}, shows the
particular wedge that connects smoothly with that for the harmonic
oscillator.

The critical case, where the upper edge of the wedge coincides
with the real axis is the case $N=4$, i.e. the potential $-x^4$.
For $N<4$, it is possible to stay on the real axis, where the wave
function decays exponentially, albeit with an oscillatory
modulation, but for $N\ge 4$ we have no option but to formulate
the problem on a contour in the lower half $x$ plane.
\begin{figure}[h]\vspace{-4cm}
\hspace{2cm}\resizebox{!}{4in}{\includegraphics{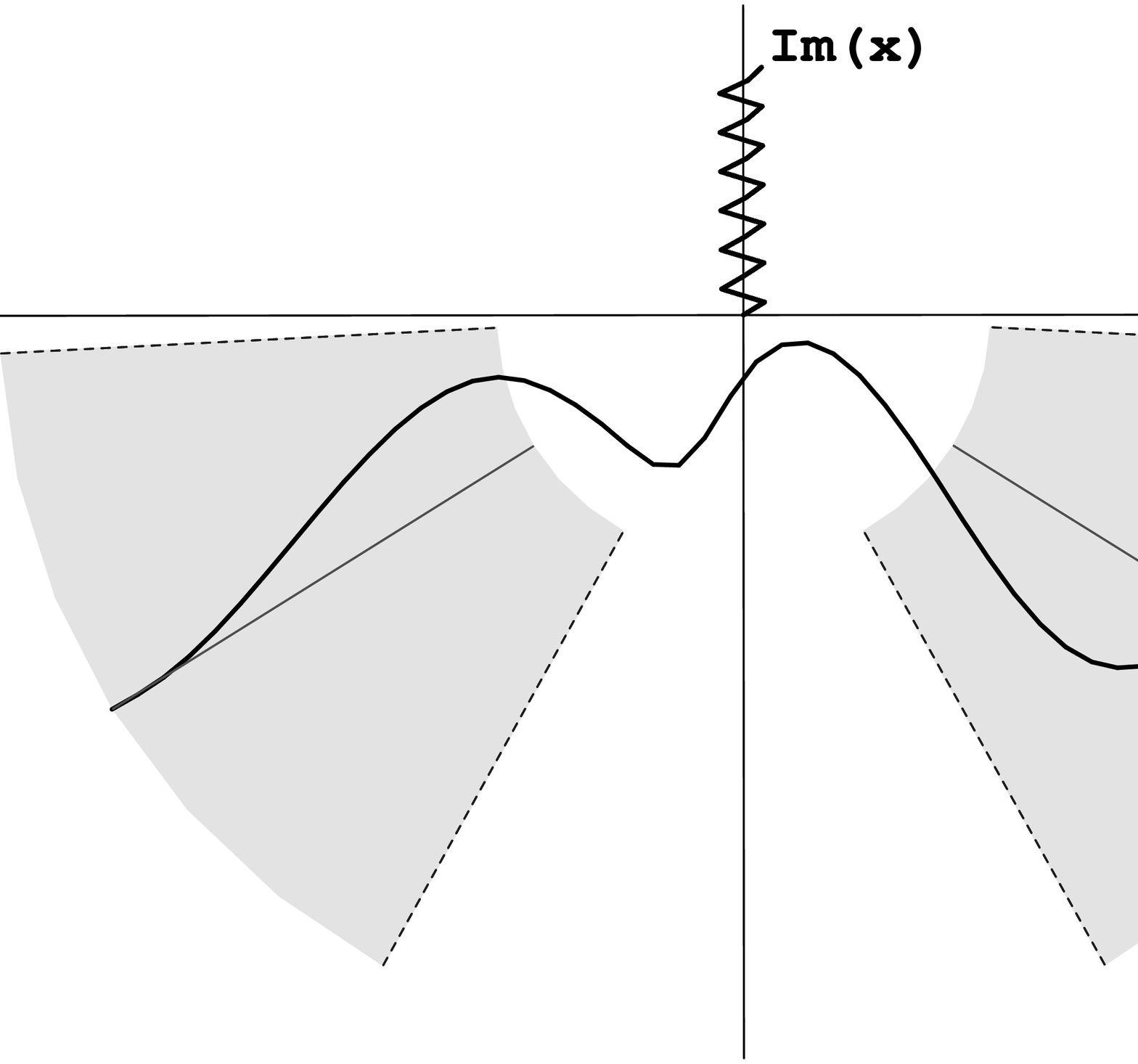}}
\vspace{-1cm}\caption{Wedges in the complex plane in which the Schr\"odinger
equation for Eq.~(\ref{HBB}) is posed.}
\end{figure}

This is the fundamental reason why the $-gx^4$ potential has
proved so intractable. At first sight it appears Hermitian: it is
only because of the contour on which it must be defined that it is
non-Hermitian. The problem is inherently non-perturbative, so any
expansion to be attempted can not be in the coupling constant $g$.
A previous attempt~\cite{WKB} used WKB methods, and was able to
calculate $Q$ to leading non-perturbative order.

\section{Choice of Contour}
Our present approach starts with the idea of
Mostafazadeh~\cite{AMR}, to map the problem back onto the real
axis using a real parametrization of a suitable contour. A wide
variety of contours are possible, as long as they go off to
infinity at an angle within the wedges. Taking $N=4$ and writing
the original $x$ variable of Eq.~(\ref{HBB}) as $z$ to reflect its
complex character, so that \beq\label{HBBz}
H=-\frac{d^2}{dz^2}-g\,z^4, \eeq the parametrization used in
Ref.~\cite{AMR} was \beq z=x \cos\theta-i|x|\sin\theta\,, \eeq
corresponding to straight-line contours, with an infinitesimal
rounding off near the origin. Here $\theta$ was taken as $\pi/6$,
the optimal angle for $N=4$. The resulting Hamiltonian  was \beq
H=e^{i{\rm sgn}(x)\pi/3}\Big(-\frac{d^2}{dx^2}+g|x|^4\Big). \eeq

Because of rounding, there are non-trivial boundary conditions at
$x=0$, namely (i) $\psi$  is real and continuous, (ii)
$\psi'(0^-)=e^{4i\theta}\psi'(0^+)$. Consequently, in calculating
$H^\dag$ there is an additional term $\delta H$ beyond the obvious
one.

A calculation of $Q$ with this Hamiltonian is very difficult
because of the boundary conditions at $x=0$ and the lack of an
obvious expansion parameter. In Ref.~\cite{HFJ+JM} we attempted to
make an expansion in $\theta$, freeing it from its
optimal value for $N=4$, noting that any positive value for
$\theta$ would suffice to make the wave function vanish with an
exponential component. In addition we smoothed out the curve
chosen in Ref.~\cite{AMR}, taking the hyperbola
\beq\label{fpath}
z=x\cos\theta-i\sin\theta\ \surd(1+x^2),
\eeq
in order to remove
the boundary conditions at $x=0$. Unfortunately this calculation
did not produce a very useful $h$, but rather one which still had
a $-x^4$ term, so that the asymptotic behaviour of the wave
function was oscillatory, with only a power suppression.

In the present paper we adopt a different approach. First we
choose a new parametrization, whose asymptotes are not in
fact in the centre of the wedges, but rather are inclined at
$\pi/4$ to the real axis,  and then we introduce an artificial
parameter $\ve$ multiplying $H_1$, the non-Hermitian part of $H(x)$.

The contour that turns out to give particularly simple results
is of the form
\beq\label{eq:path}
z=-2i\surd(1+i\,x).
\eeq
Notice that with this choice, the $PT$-symmetry of the original Hamiltonian,
which is a real function of $iz$, will be respected by the new Hamiltonian,
written in terms of $x$. This new Hamiltonian is in fact
\beq
  \label{eq:ham}
  H=\half\{(1+i\,x),p^{2}\}-\half p-\cop(1+i\,x)^2,
\eeq
where $\{..\,,..\}$ denotes the anticommutator,  $p\equiv d/dx$, and for
convenience we have introduced $\cop\equiv 16 g$.
Separating $H$ into its Hermitian and
anti-Hermitian parts, and multiplying the latter by the artificial
parameter $\ve$, which at the end will be set equal to one, we write
\beq
  \label{eq:expan}
  H=H_0+\ve\,H_1,
\eeq
where
\bea
  \label{eq:ham12}
  H_{0}&=& p^2-\half p+\cop(x^2-1)\nonumber\\
  H_{1}&=&\half\, i\,\{x,p^{2}\}-2i\cop\,x
\eea
\section{Calculation of $Q$ and $h$}

First we calculate $Q_1$ from the first of Eqs.~(\ref{Qs}), namely $[Q_1,H_0]=2H_1$.
As a general, systematic procedure for such problems we would write the Hermitian
operator $Q_{1}$ as a sum of anticommutators of the form
$Q_{1}=\sum_{n\,\textrm{odd}}\{f_{n}(x),p^{n}\}$, where $f_{n}(x)$ is a real function of $x$,
and gradually increase the order $n$.  However, in this case $H_0$ and $H_1$ are so simple
that the solution can essentially be found by inspection. Thus a $p^3$ term in $Q_1$
will produce the desired structure $i\,\{x,p^{2}\}$ when commuted with the $x^2$ term of $H_0$,
while a term in $p$ will produce the $x$ term of $H_1$. By equating coefficients we find that
\beq
  \label{eq:q1}
  Q_1=-\frac{p^3}{3\cop}+2p\,.
\eeq
In order to calculate $Q_3$ from the second of Eqs.~(\ref{Qs}) we need the double commutator
$[Q_1,[Q_1,H_1]]$. First let us calculate the inner commutator $[Q_1,H_1]$, which will also
be needed for the computation of $h$:
\beq\label{eq:comq1h1}
[Q_1,H_1]=-\frac{p^4}{\cop}+4p^2-4\cop.
\eeq
The crucial point is that this is a function of $p$ only, and therefore commutes with $Q_1$.
Thus $[Q_1,[Q_1,H_1]]=0$, which means that $Q_3=0$. Then the third of Eqs.~(\ref{Qs}) shows that
$Q_5=0$ and so on. Thus we have an {\it exact} solution for $Q$, after setting $\ve=1$, namely
\beq\label{eq:Q}
\mbox{\fbox{$\;Q=-\begin{array}{c}\underline{\,p^3}\\ 3\cop\end{array}+2p\;$}}
\eeq
Having obtained the metric operator $Q$ we are in a position to calculate
the equivalent Hermitian Hamiltonian $h$ of Eq.~(\ref{h}). Because the expansion for $Q$
has truncated, so does that for $h$, namely $h=H_0+h_2$. The commutator required for the evaluation
of $h_2$ has already been calculated in Eq.~(\ref{eq:comq1h1}), so it is straightforward to
evaluate $h$, with the remarkably simple result that
\beq\label{eq:h}
h=\frac{p^4}{4\cop}-\half p +\cop x^2.
\eeq
We emphasize that this Hermitian Hamiltonian, defined on the real line,
has the same energy spectrum as that of the original $H$ of Eq.~(\ref{HBBz}) defined
on a complex contour. The only unusual feature of $h$ is that it does not have the standard form
of a quadratic kinetic term plus a potential. However, just such a Hamiltonian results if
we take the Fourier transform. In terms of the transformed variable $y$, and after a rescaling $y\to y\sqrt{\cop}$, we have
\beq
  \label{eq:fortran}
\mbox{\fbox{$\;\tilde{h}= p^2_{y}+\begin{array}{c}\underline{\,1\,}\rule{0cm}{.5cm}\\4\end{array}\cop y^4
-\begin{array}{c}\underline{\,1\,}\rule{0cm}{.5cm}\\2\end{array}\sqrt{\cop}\, y\;$}}
\end{equation}
\section{Discussion}
Equations (\ref{eq:Q}) and (\ref{eq:fortran}) constitute our main results.
The latter exhibits a standard Hermitian Hamiltonian, with a
positive quartic potential plus a linear term, shown in Fig.~3, whose spectrum is the same as that of the original
problem, with a
$-z^4$ potential posed on a contour in the complex plane. It constitutes the first direct, constructive
proof of the reality of the spectrum of Eq.~(\ref{HBBz}). In accordance with our introductory remarks, we
note that $\tilde{h}$ is completely non-perturbative, since, without a harmonic term $m^2x^2$ term in the potential,
$g$ can rescaled to 1.
\begin{figure}[h]
 \centerline{\resizebox{!}{3.8in}{\includegraphics{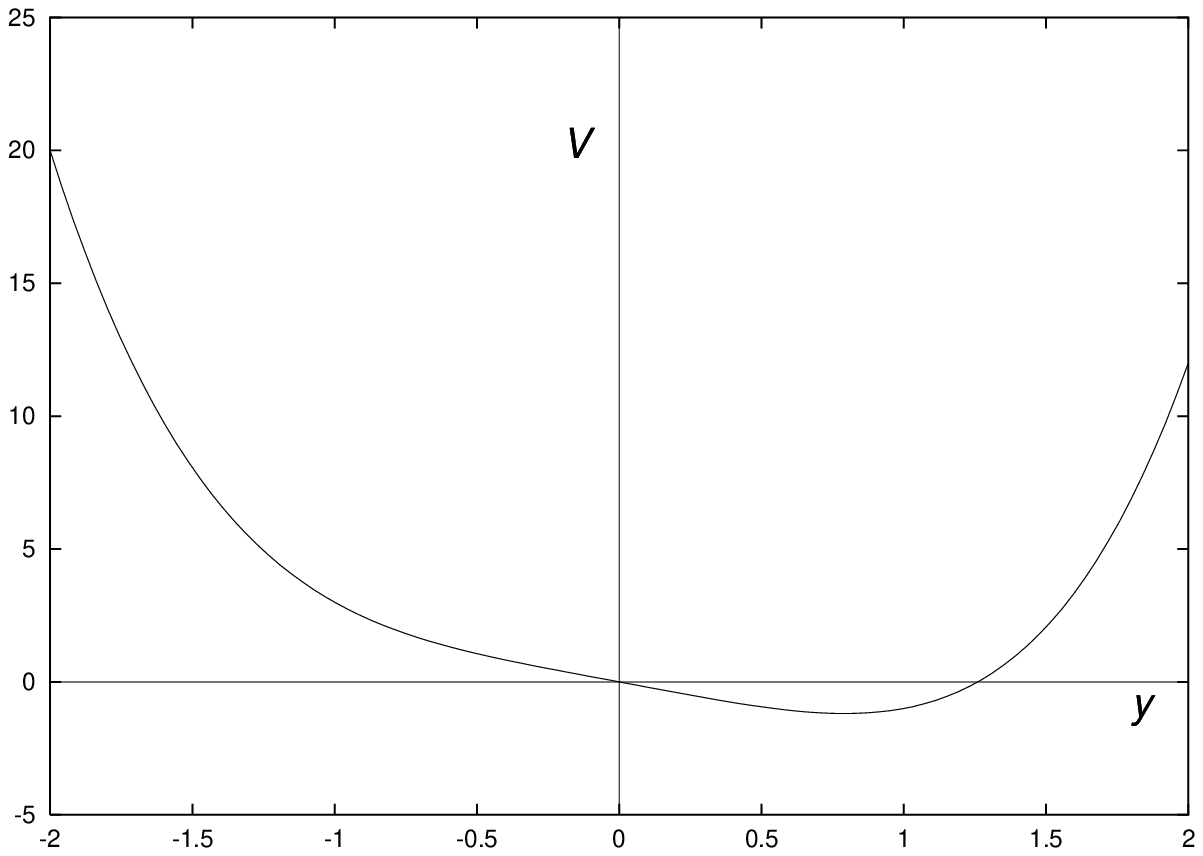}}}
 \caption{The potential of Eq.~(\ref{eq:fortran}), with $\cop=16$ ($g=1$).}
 \end{figure}

We have performed a numerical calculation of the energy eigenvalues of Eq.~(\ref{eq:fortran}),
using both Runge-Kutta integration and the
variational truncated matrix method of Ref.~\cite{Paolo}. Both
methods give eigenvalues that are indistinguishable from those
cited by Bender and Boettcher (calculated by Runge-Kutta
integration along a complex contour) in their original
paper~\cite{BB}.

A simple extension of the above result can be obtained when an additional harmonic term $m^2z^2$
is introduced into Eq.~(\ref{HBBz}). The only change in Eq.~(\ref{eq:h}) is
that $h$ becomes
\beq
h=\frac{(p^2-4m^2)^2}{4\cop}-\half p +\cop x^2
\eeq
with corresponding scaled Fourier transform
\beq\label{BG}
\tilde{h}=p_y^2+\frac{1}{4\cop}(\cop y^2-4m^2)^2-\half\sqrt{\cop}\,y
\eeq
After completion of this work we were made aware of an earlier paper by Buslaev and Grecchi~\cite{BG},
which showed the spectral equivalence of the massive version of the $-x^4$ theory (their $H_\ve(ig,j)$,
with $j=1$), formulated on the line $z=x-i\eta$, with a Hermitian Hamiltonian that can be identified with
Eq.~(\ref{BG}) on setting $\cop=4g^2$, $m=1/2$. Their method made use of the perturbation
series for the energy eigenvalues of the two Hamiltonians, which only exists for $m\ne 0$.
However, they were subsequently able to go the massless limit by rescaling and taking $g$ to $\infty$.
In this way they obtained the spectral equivalence between Eq.~(\ref{HBBz}) and Eq.~(\ref{eq:h}) (see their
Theorem 6, with $j=1$, $\alpha=0$ after a simple rescaling).

The present paper approaches the problem from a completely different perspective and offers
a simple, explicit and transparent derivation of these spectral equivalences, together with the operator
$Q$ required to define the positive-definite metric, and the
observables~\cite{AM}, of the non-Hermitian Hamiltonians.\\

{\bf Acknowledgements}

This research has been supported in part by JCyL under a Research Grant,
by MEC under contract MTM 2005-09183 and by JCyL under contract VA013C05.
We are grateful to C.~Bender, P.~Dorey, L.~M.~Nieto and J.~Negro for useful discussions,
to P.~Dorey and M.~Znojil for bringing Ref.~\cite{BG} to our attention, and to
D.~Brody for a careful reading of the manuscript.

\end{document}